\documentclass[useAMS,usenatbib]{mn2e}
\usepackage{txfonts}
\usepackage{natbib}
\usepackage[all]{xy}
\usepackage[dvips]{graphicx}
\usepackage{epstopdf}

\def\del#1{{}}

\newcommand{\ltsima}{$\; \buildrel < \over \sim \;$}
\newcommand{\lsim}{\lower.5ex\hbox{\ltsima}}
\newcommand{\gtsima}{$\; \buildrel > \over \sim \;$}
\newcommand{\gsim}{\lower.5ex\hbox{\gtsima}}
\newcommand{\bra}{\langle}
\newcommand{\ket}{\rangle}

\newcommand{\dd}{\mathrm{d}}

\newcommand{\ci}{\mathrm{i}}

\newcommand{\trace}{\mathrm{tr}}

\newcommand{\lprime}{\ell^\prime}
\newcommand{\dirac}{\delta_D}

\newcommand{\vecx}{\bmath{x}}
\newcommand{\veck}{\bmath{k}}
\newcommand{\veca}{\bmath{\alpha}}
\newcommand{\vect}{\bmath{\theta}}

\newcommand{\vecl}{\bmath{L}}
\newcommand{\lhat}{\hat{L}}

\title[Evolution of ellipticity correlations]
{Evolution of intrinsic ellipticity correlations due to peculiar motion}
\author[A. Giahi-Saravani and B.M. Sch{\"a}fer]{
Aram Giahi-Saravani\thanks{aram@ari.uni-heidelberg.de} and Bj{\"o}rn Malte Sch{\"a}fer\\
Astronomisches Recheninstitut, Zentrum f{\"u}r Astronomie, Universit{\"a}t Heidelberg, M{\"o}nchhofstra{\ss}e 12, 69120 Heidelberg, Germany}

\begin{document}
\pagerange{\pageref{firstpage}--\pageref{lastpage}}
\pubyear{2008}
\maketitle
\label{firstpage}

\begin{abstract}
Topic of this paper is the time-evolution of intrinsic correlations of galaxy ellipticities due to peculiar motion. In our model, the galaxy ellipticities are determined from the angular momentum of their host haloes, which can be computed from the fluctuations statistics of a Gaussian random field. Subsequent peculiar motion distorts the ellipticity field and causes changes in the ellipticity correlations. Using analogies between this problem of shifted ellipticity tensors and the displacements of polarisation tensors in gravitational lensing of the cosmic microwave background we compute $E$-mode and $B$-mode spectra of the time-evolved ellipticity field, where the displacements are modelled with first and second order Lagrangian perturbation theory. For EUCLID, ellipticity correlations are decreased on large multipoles $\ell\gsim1000$, amounting to up to 10\% in the $E$-mode spectrum $C_E^\epsilon(\ell)$ and up to 60\% in the $B$-mode spectrum $C_B^\epsilon(\ell)$ at $\ell\simeq3000$ due to the dispersing effect of peculiar motion. $E/B$-mode conversion in analogy to CMB-lensing is present but small. We conclude that distortions of the ellipticity field due to peculiar motion is not affecting the prediction of ellipticity models on the scales relevant for lensing in the case of EUCLID's galaxy distribution, but should affect larger scales for surveys at lower redshifts.
\end{abstract}


\section{Introduction}
Weak gravitational lensing by the cosmic-large scale structure is a tool for investigating the fluctuations statistics of the cosmic density field and its dependence on the underlying cosmological model \citep[for reviews, see][]{2001PhR...340..291B, 2010CQGra..27w3001B}. Future weak lensing surveys such as EUCLID, DES, LSST and JDEM  are designed to yield sub-percent accuracy on the set of cosmological parameters in a dark energy cosmology by measuring the correlation function of the gravitationally sheared galaxy ellipticity field. A common assumption in weak cosmic lensing is that those ellipticities are intrinsically uncorrelated and that the only correlating effect is weak lensing, because the light from neighbouring galaxies has to transverse the same cosmic tidal fields, leading to a correlation in change in shape. 

This assumption, however, is challenged on small scales by intrinsic alignment effects \citep[for a review on angular momentum models and intrinsic alignments, see][]{2009IJMPD..18..173S}. Tidal shearing models for angular momentum built-up in galactic haloes predict correlated angular momenta of neighbouring galaxies. If the symmetry axis of the galactic disk is aligned with the angular momentum direction of the host halo, neighbouring galaxies are viewed under correlated angles of inclination such that their ellipticities appear correlated. This intrinsic alignment effect is important on small scales because the angular momentum correlation is comparatively short ranged: it is predicted to be present on scales of about $1~\mathrm{Mpc}/h$ \citep{2001ApJ...559..552C, 2001PASA...18..198N, 2011arXiv1101.4584S}.

Intrinsic alignments of galaxies based on angular momentum models are a relatively new topic which will undoubtedly attract much interest in the future as the weak lensing data sets provided by large-scale lensing observations will at the same time help to scrutinise intrinsic alignment models. The theory of angular momentum-induced alignments \citep{2000ApJ...545..561C, 2001ApJ...559..552C,2002ApJ...568...20C, 2002MNRAS.332..788M} has been applied to describe contamination of weak lensing data in the convergence spectrum \citep{2000MNRAS.319..649H,2003MNRAS.339..711H,2004MNRAS.347..895H, 2006MNRAS.371..750H} and bispectrum \citep{2008arXiv0802.3978S}. Different schemes for removing the contamination of intrinsic alignments have been proposed, from discarding close galaxy pairs \citep{2003A&A...398...23K,2002A&A...396..411K} to specifically designed weighting schemes for nulling out their contribution \citep{2008arXiv0804.2292J} or amplifying them relative to the weak lensing induced ellipticity correlations \citep{2003A&A...398...23K, 2010A&A...517A...4J}. Resulting biases on cosmological parameter estimation if intrinsic alignments remain uncorrected have been quantified \citep{2007NJPh....9..444B, 2010A&A...523A...1J, 2010MNRAS.408.1502K, 2010MNRAS.402.2127S}.

The wealth of structure in the angular momentum and ellipticity field and their alignment with large-scale tidal fields has attracted much interest from the numerical perspective \citep{2007MNRAS.381...41H,2012arXiv1201.5794C} and suggests the question if large-scale tidal fields can be reconstructed using the ellipticity field as a tracer \citep{2000ApJ...532L...5L, 2001ApJ...555..106L, 2007arXiv0707.1690L}.

On a more fundamental level, the investigation of tidal shearing mechanisms in different orders of perturbation theory along with the deformation of forming haloes due to tidal forces \citep{1996MNRAS.282..455C,1995MNRAS.276..115C,2001MNRAS.320L...7C,1996MNRAS.282..436C,1995MNRAS.276...39C,2001MNRAS.323..713C, 2007arXiv0709.1106L}. Numerical verification of the tidal torquing theories of angular momentum build-up has been the topic of a number of papers \citep{2001MNRAS.323..713C, 2001ApJ...555..240B, 2002MNRAS.332..325P, 2002MNRAS.332..339P, 2007MNRAS.375..489H} who agree that the angular momentum direction can be well described by tidal torquing whereas the amount of angular momentum might be overestimated. Because ellipticity alignments are only sensitive to the angular momentum direction, these studies provide support for using tidal torquing models with this particular application in mind.

The way in which the orientation of a galactic disk is linked to the angular momentum direction of the host halo is not very clearly cut. In a small number of observations \citet{2005ApJ...627L..17B} found mismatches and suggest that direct linking of the symmetry axis of the disk to the host halo angular momentum would lead to overestimation of the ellipticity alignments. This effect is partially covered by  parameterisation, but unless the relation is better understood, angular momentum-based alignment models proved upper limits on ellipticity correlations. By now, intrinsic alignments have been measured in a number of data sets and have been found at the expected levels \citep{2000ApJ...543L.107P, 2002ApJ...567L.111L, 2006MNRAS.367..611M,2007MNRAS.381.1197H}, although some studies doubt these claims \citep[e.g.][]{2011MNRAS.tmp.1665A}.

The point which motivated this paper is the comparatively short-ranged correlation of angular momenta and consequently of the galaxy ellipticities, which reaches out to distances of about $1~\mathrm{Mpc}/h$. If the ellipticity field is this short-ranged, and if it is distorted by the peculiar motion of galaxies, by how much do the correlations change and on what scales? We will investigate this question by employing a formalism based on lensing of the CMB-polarisation, by describing peculiar motion with Lagrangian perturbation theory, and consider the EUCLID galaxy sample as an application: After a summary of cosmology, structure formation, Lagrangian perturbation theory, angular momentum models and ellipticity correlations in Sect.~\ref{sect_cosmology} we describe our formalism and the results in Sect.~\ref{sect_evolution}. Our main findings are summarised in Sect.~\ref{sect_summary}.

As reference model we chose a spatially flat $w$CDM model with Gaussian adiabatic initial perturbations in the cold dark matter distribution. Specifically, parameters were chosen to be $\Omega_m = 0.25$, $n_s = 1$, $\sigma_8 = 0.8$, $\Omega_b=0.04$ and finally $H_0=10^5\: h\:\mathrm{m}/\mathrm{s}/\mathrm{Mpc}$, with $h=0.72$. The dark energy equation of state is set to $w=-0.95$ and the sound speed is equal to the speed of light, $c_s=c$, such that there is no dark energy clustering.

\section{cosmology}\label{sect_cosmology}

\subsection{Dark energy cosmologies}
The dynamics of a spatially flat Friedmann-universe with dark matter and dark energy is described by the Hubble function $H(a)=\dd\ln a/\dd t$, which is given by
\begin{equation}
\frac{H^2(a)}{H_0^2} = \frac{\Omega_m}{a^{3}} + (1-\Omega_m)\exp\left(3\int_a^1\dd\ln a\:(1+w(a))\right),
\end{equation}
with the matter density parameter $\Omega_m$ and the dark energy equation of state function $w(a)$. The value $w\equiv -1$ corresponds to the cosmological constant $\Lambda$. Comoving distance $\chi$ and scale factor $a$ are related by
\begin{equation}
\chi = c\int_a^1\:\frac{\dd a}{a^2 H(a)},
\end{equation}
which yields distances in unit of the Hubble distance $\chi_H=c/H_0$. For the galaxy redshift distribution $n(z)\dd z$, we use a standard shape
\begin{equation}
n(z) = n_0\left(\frac{z}{z_0}\right)^2\exp\left(-\left(\frac{z}{z_0}\right)^\beta\right)\dd z
\quad\mathrm{with}\quad \frac{1}{n_0}=\frac{z_0}{\beta}\Gamma\left(\frac{3}{\beta}\right),
\end{equation}
with parameters $z_0=0.64$ and $\beta=3/2$, as forecasted for EUCLID \citep{2007MNRAS.381.1018A}. The distribution can be rewritten in terms of comoving distance using the relation $p(z)\dd z = p(\chi)\dd\chi$ with $\dd z/\dd\chi = H(\chi)/c$.

\subsection{CDM power spectrum}
The spectrum $P(k)$ describes the fluctuation amplitude of the Gaussian, statistically homogeneous density field $\delta$, $\bra\delta(\bmath{k})\delta(\bmath{k}^\prime)\ket=(2\pi)^3\dirac(\bmath{k}+\bmath{k}^\prime)P(k)$, and is given by the ansatz
\begin{equation}
P(k)\propto k^{n_s}T^2(k),
\end{equation}
with the transfer function $T(k)$. This transfer function is approximated by
\begin{equation}
T(q) = \frac{\ln(1+2.34q)}{2.34q}\left(1+3.89q+(16.1q)^2+(5.46q)^3+(6.71q)^4\right)^{-\frac{1}{4}},
\label{eqn_cdm_transfer}
\end{equation}
\citep[see][]{1986ApJ...304...15B}. The wave vector $k=q\Gamma$ is measured in units of the shape parameter $\Gamma$. \citet{1995ApJS..100..281S} describe corrections due to a non-zero baryon density $\Omega_b$,
\begin{equation}
\Gamma=\Omega_m h\exp\left(-\Omega_b\left[1+\frac{\sqrt{2h}}{\Omega_m}\right]\right).
\end{equation}
The spectrum $P(k)$ is normalised to the variance $\sigma_8$ on the scale $R=8~\mathrm{Mpc}/h$,
\begin{equation}
\sigma^2_R = \int\frac{k^2\dd k}{2\pi^2}\: P(k) W^2(kR)
\end{equation}
with a Fourier transformed spherical top hat filter function, $W(x)=3j_1(x)/x$. $j_\ell(x)$ is the spherical Bessel function of the first kind of order $\ell$ \citep{1972hmf..book.....A}.

\subsection{Structure growth with clustering dark energy}
The growth of the density field in the linear regime, $\delta(\bmath{x},a)=D_+(a)\delta(\bmath{x},a=1)$, is given by the growth function $D_+(a)$, which follows as a solution to the growth equation \citep{1997PhRvD..56.4439T, 1998ApJ...508..483W, 2003MNRAS.346..573L},
\begin{equation}
\frac{\dd^2}{\dd a^2}D_+(a) + \frac{1}{a}\left(3+\frac{\dd\ln H}{\dd\ln a}\right)\frac{\dd}{\dd a}D_+(a) = 
\frac{3}{2a^2}\Omega_m(a) D_+(a).
\label{eqn_growth}
\end{equation}

\subsection{Lagrangian perturbation theory}
The peculiar motion of galaxies can be described using Lagrangian perturbation theory (LPT) if the flow of dark matter and of the advected galaxies is irrotational and nonlinearities are weak. In this limit, galaxies follow straight lines given by the gradient of the Zel'dovich potential $\Phi_1$ to first order \citep[1LPT,][]{1970A&A.....5...84Z, 1970Afz.....6..581D, 1989A&A...223....9B, 1991ApJ...382..377M, 2002PhR...367....1B},
\begin{equation}
\vecx\rightarrow\vecx - D_1(a)\nabla\Phi_1
\end{equation}
where $\Phi_1$ is the solution to the Poisson equation, $\Delta\Phi_1 = \Delta\Phi = \delta$. This solution can be improved by adding second order corrections to Lagrangian perturbation theory \citep[2LPT,][]{1994MNRAS.267..811B, 1995A&A...294..345M, 1995A&A...296..575B},
\begin{equation}
\vecx\rightarrow\vecx - D_1(a)\nabla\Phi_1 + D_2(a)\nabla\Phi_2
\end{equation}
with the second order potential $\Phi_2$ \citep{1994A&A...288..349B, 1995A&A...296..575B}, 
\begin{equation}
\Delta\Phi_2 = \sum_{i>j}\left[\Phi_{ii}\Phi_{jj} - \Phi_{ij}\Phi_{ij}\right].
\end{equation}
The time dependences are given by $D_1(a)=D_+(a)$ and $D_2(a) = -3/7 D_+^2(a)\Omega_m^{-1/143}$ \citep[for a low $\Omega_m$-cosmology with a cosmological constant $\Lambda$, see][]{1992ApJ...394L...5B}. The solution to the latter relation can be written down in Fourier-space, where the products of tidal fields become convolutions,
\begin{equation}
\Phi_2 = -\frac{1}{k^2}\int\frac{\dd^3k^\prime}{(2\pi)^3}
\sum_{i>j} Q_{ij}(\veck,\veck^\prime)\: \delta(\veck^\prime)\delta(\veck-\veck^\prime),
\end{equation}
where the mode coupling function $Q_{ij}(\veck,\veck^\prime)$ becomes:
\begin{equation}
Q_{ij}(\veck,\veck^\prime) = \frac{(\veck^\prime)_i^2(\veck^\prime-\veck)_j^2 - \veck_i\veck_j(\veck-\veck^\prime)_i(\veck-\veck^\prime)_j}{(\veck^\prime)^2(\veck-\veck^\prime)^2}.
\end{equation}
Spectra of the potentials $\Phi_1$ and $\Phi_2$ can be defined by
\begin{equation}
\bra\Phi_i(\veck)\Phi_i(\veck^\prime)\ket = (2\pi)^3\delta_D(\veck+\veck^\prime) P_\Phi^{(i)}(k),
\quad i=1,2,
\end{equation}
with $P_\Phi^{(1)}(k) = P(k)/k^4$ as a consequence of the Poisson equation and with $P_\Phi^{(2)}(k)$ which can be derived to follow
\begin{equation}
P_\Phi^{(2)}(k) = \frac{2}{k^4}\int\frac{\dd^3 k^\prime}{(2\pi)^3}\:
\left(\sum_{i>j} Q_{ij}(\veck^\prime,\veck-\veck^\prime)\right)^2
P(\left|\veck^\prime\right|) P(\left|\veck-\veck^\prime\right|)
\end{equation}
by application of the Wick-theorem \citep[for a proof, see][]{2008cmbg.book.....D}. The integration is most efficiently carried out using cylindrical coordinates aligned with $\veck$ such that $\dd^3k^\prime = 2\pi (k^\prime)^2\dd k^\prime\dd\cos\theta$ using azimuthal symmetry, with $\theta$ being the angle between $\veck$ and $\veck^\prime$.

\begin{figure}
\begin{center}
\resizebox{\hsize}{!}{\includegraphics{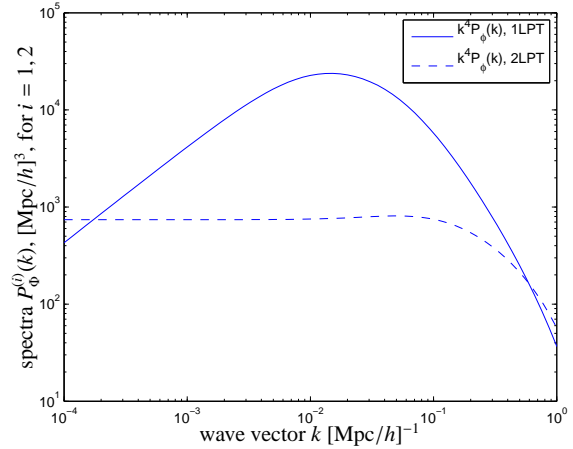}}
\end{center}
\caption{CDM spectra $k^4P_\Phi^{(i)}(k)$, $i=1,2$ which are employed in Lagrangian perturbation theory for displacing the galaxies, for 1LPT (solid line) and 2LPT (dashed line). The 1LPT result corresponds in this representation to the CDM spectrum $P(k)$: $k^4 P_\Phi^{(1)}(k) = P(k)$}
\label{fig_potential}
\end{figure}

Fig.~\ref{fig_potential} gives an impression of the spectrum $P_\Phi^{(1)}(k)$ and of the 2LPT-corrections $P_\Phi^{(2)}(k)$ relative to 1LPT. We plot $k^4P_\Phi^{(i)}(k)$, $i=1,2$ which is equal to the CDM spectrum $P(k)$ for the 1LPT result due to the Poisson equation. The 2LPT-spectrum is smaller on almost all scales by up to an order of magnitude and is only similar in amplitude on spatial scales of about $1~\mathrm{Mpc}/h$. 

\subsection{Angular momentum from tidal shearing}
Angular momenta of dark matter haloes are introduced by tidal shearing, where the differential motion of a protohalo gives rise to a torquing moment \citep{1949MNRAS.109..365H, 1955MNRAS.115....2S, 1969ApJ...155..393P, 1970Afz.....6..581D, 1984ApJ...286...38W}:
\begin{equation}
L_\alpha = a^3 H(a)\frac{\dd D_+}{\dd a}\epsilon_{\alpha\beta\gamma}\sum_\delta I_{\beta\delta}\Phi_{\delta\gamma}
\end{equation}
This relation reflects the interesting misalignment property of the shear and inertia eigensystems necessary for angular momentum generation \citep{2011arXiv1101.4584S}: Only the antisymmetric tensor $X^-_{\beta\gamma} = \sum_\delta (I_{\beta\delta}\Phi_{\delta\gamma} - \Phi_{\beta\delta}I_{\delta\gamma})/2$ is relevant for the angular momentum, $L_\alpha\propto X^-_{\beta\gamma}$, because the contraction of the symmetric tensor $X^+_{\beta\gamma} = \sum_\delta (I_{\beta\delta}\Phi_{\delta\gamma} + \Phi_{\beta\delta}I_{\delta\gamma})/2$ with the antisymmetric $\epsilon_{\alpha\beta\delta}$ vanishes. The antisymmetric tensor $X_-$ is equal to the commutator $[I_{\beta\delta}, \Phi_{\delta\gamma}]$ which suggests that for angular momentum generation, the tidal shear and the inertia are not allowed to be simultaneously diagonalisable and may not have a common eigensystem.

Angular momenta $\vecl$ are described as being coupled to the tidal shear by means of a Gaussian random process $p(\vecl|\Phi_{ij})\dd\vecl$ involving tidal fields $\Phi_{ij}$ shaping the covariance $\mathrm{cov}(L)_{ij}$ of the Gaussian distribution \citep{2001ApJ...555..106L},
\begin{equation}
\mathrm{cov}(L)_{ij} = \bra L_i L_j\ket = \frac{\bra\vecl^2\ket}{3}\left(\frac{1+a}{3}\delta_{ij} - a\: (\hat\Phi^2)_{ij}\right),
\end{equation}
with the misalignment parameter $a$, which describes the average orientation of the protohalo's inertia to the tidal shear eigensystem. $a$ has been measured in numerical simulation to be close to 0.25 which we will assume in this work. $\hat{\Phi}$ is the unit normalised traceless tidal shear with the properties $\trace(\hat\Phi)=0$ and $\trace(\hat\Phi^2)=1$. This description is valid on scales where the correlations between tidal shears are negligible.

\subsection{Intrinsic ellipticity correlations}
Ellipticity correlations between galaxies are traced back to correlated angular momenta of their host haloes. CDM haloes acquire their angular momentum by tidal shearing and due to the fact that neighbouring galaxies experience correlated tidal fields, their angular momenta are correlated in consequence. The angular momentum $\vecl$ in turn determines the angle of inclination at which the galactic disk is viewed, and ultimately the ellipticity $\epsilon$ \citep{2000MNRAS.319..649H, 2001ApJ...559..552C, 2002ApJ...568...20C, 2002MNRAS.332..788M, 2003MNRAS.339..711H}:
\begin{equation}
\epsilon=\epsilon_+ +\ci\epsilon_\times
\quad\mathrm{with}\quad
\epsilon_+ = \alpha\frac{\lhat_x^2-\lhat_y^2}{1+\lhat_z^2},\quad
\epsilon_\times = 2\alpha\frac{\lhat_x\lhat_y}{1+\lhat_z^2},
\end{equation}
with the angular momentum direction $\lhat = \vecl/L$ and the coordinate system being aligned with its $z$-axis being parallel to the line of sight. A rotation of the coordinate frame by $\varphi$ causes the complex ellipticity to rotate twice as fast, $\epsilon\rightarrow\exp(2\ci\varphi)\epsilon$, in accordance with the spin-2 property of the ellipticity field. $\alpha$ is a free parameter weakening the dependence between inclination angle and ellipticity for thick galactic disks and has been determined to be $\alpha = 0.75$ in the APM sample \citep{2001ApJ...559..552C}.

In this work we use the angular momentum-based ellipticity correlation model proposed by \citet{2001ApJ...559..552C}, who trace ellipticity correlations back to tidal shear correlations using the conditional probability distribution $p(\vecl|\Phi_{ij})\dd\vecl$ introduced by \citet{2001ApJ...555..106L}: In this model, the distribution $p(\vecl|\Phi_{ij})\dd\vecl$ is assumed as being Gaussian which is then being marginalised over the magnitude of the angular momentum vector, retaining only its directional dependence. Writing down the ellipticity components as a function of the angular momentum direction and employing the covariance $\bra L_i L_j\ket$ as a function of the squared tidal shear tensor, as advocated by Lee and Pen, it is possible to relate the tidal shear correlations to the spectrum of the density field. With this relation, one can write down a correlation function of the ellipticity field as a function of moments $\zeta_n(r)$ \citep[see][]{2001ApJ...559..552C} of the tidal shear field and finally to carry out a Limber projection for obtaining the angular correlation function. For the parameter $a$ we chose the value 0.25 supported by numerical simulations.

Ellipticity correlations between two points $\vect_1$ and $\vect_2$ separated by the distance $\theta$ are described in terms of two correlation functions $\xi_\pm(\theta)$,
\begin{eqnarray}
\xi_+(\theta) & = & \bra\epsilon^*(\vect_1)\epsilon(\vect_2)\ket =
\bra \epsilon_+(\vect_1)\epsilon_+(\vect_2)\ket + 
\bra\epsilon_\times(\vect_1)\epsilon_\times(\vect_2)\ket \\
\xi_-(\theta) & = & \bra\epsilon(\vect_1)\epsilon(\vect_2)\ket =
\bra \epsilon_+(\vect_1)\epsilon_+(\vect_2)\ket - 
\bra\epsilon_\times(\vect_1)\epsilon_\times(\vect_2)\ket
\end{eqnarray}
which are formed from the variances of the ellipticity components $\epsilon_+$ and $\epsilon_\times$ using $\bra\epsilon_+\epsilon_\times\ket=0$. They can be transformed to the spectra $C_E^\epsilon(\ell)$ and $C_B^\epsilon(\ell)$ of the gradient and vorticity modes of the ellipticity field,
\begin{eqnarray}
C_E^\epsilon(\ell) & = & \pi\int\theta\dd\theta 
\left[\xi_+(\theta)J_0(\ell\theta) + \xi_-(\theta)J_4(\ell\theta)\right],
\label{eqn_e_transform}\\
C_B^\epsilon(\ell) & = & \pi\int\theta\dd\theta 
\left[\xi_+(\theta)J_0(\ell\theta) - \xi_-(\theta)J_4(\ell\theta)\right],
\label{eqn_b_transform}
\end{eqnarray}
by Fourier transform \citep{1992ApJ...388..272K, 2002A&A...389..729S, 2007A&A...462..841S,2010MNRAS.401.1264F}. Fig.~\ref{fig_ellipticity} shows intrinsic ellipticity spectra $C_E^\epsilon(\ell)$ and $C_B^\epsilon(\ell)$ for the EUCLID galaxy sample with its median redshift at $z_\mathrm{med}=0.9$. The spectra are constant and equal in amplitude up to multipoles of $\ell\simeq100$, indicating the absence of correlations such that on each scale on measures the variance of the uncorrelated ellipticity field. Correlations become important on angular scales $\ell\gsim 300$ where the spectra level off and decrease from multipoles of $\ell\gsim3000$ on very rapidly. In the peak region, the ellipticity $E$-modes have an amplitude larger than the $B$-modes by about an order of magnitude.

\section{Evolution of ellipticity correlations}\label{sect_evolution}

\subsection{Analogy between ellipticities and polarisation}
The evolution of the angular ellipticity spectra due to peculiar motion of the galaxies are described in our model by drawing an analogy to lensing of the polarisation modes of the cosmic microwave background. Both the galaxy ellipticities and the Stokes-parameters of the CMB-polarisation form a tensorial spin-2 field, which means that rotations of the coordinate frame by an angle $\varphi$ give rise to a transformation of the tensor components as $\epsilon\rightarrow\exp(2\ci\varphi)\epsilon$ and $P\rightarrow \exp(2\ci\varphi) P$, when the ellipticity is written as a complex ellipticity $\epsilon = \epsilon_++\ci\epsilon_\times$ and the polarisation tensor $P$ is composed of the Stokes parameters $Q$ and $U$ according to $P = U+\ci Q$.

Peculiar motion as well as gravitational lensing introduces a shift in the position by an angle $\alpha$ such that the ellipticity $\epsilon$ is not observed at the position $\vect$ where the galaxy was formed, but rather has been displaced $\epsilon(\vect)\rightarrow\epsilon(\vect+\veca)$. The correlation properties of such a distorted field can be computed using the formalism developed for CMB lensing, which allows the computation of correlation of the lensed polarisation field, $P(\vect)\rightarrow P(\vect+\veca)$, where $\alpha$ refers now to the lensing deflection angle. Our formalism will be built in complete analogy and computes the shifting angle from the peculiar velocity, which in turn is derived from a velocity potential using Lagrangian perturbation theory for the description of peculiar motion.

\subsection{Formalism for displacing the ellipticities}
By drawing analogies between the peculiar motion of galaxies causing displacements in the ellipticities, $\epsilon(\vect)\rightarrow\epsilon(\vect+\veca)$ and the lensing of the polarisation of the CMB, $P(\vect)\rightarrow P(\vect+\veca)$ it becomes possible to derive spectra of the evolved ellipticity field. Peculiar motion by $D_+(a)\nabla\Phi$ changes the position of a galaxy by a shifting angle $\veca = D_+\nabla\Phi/\chi$ if the galaxy is situated at a comoving distance $\chi$. The angular displacement field $\veca$ can be derived from a displacement potential $\psi = D_+\Phi/\chi^2$ by angular derivation, such that $\veca = \nabla_\theta\psi$, because $\nabla_\theta = \chi\nabla$. Generalising this argument to a galaxy population which is described by a normalised distribution $n(\chi)\dd\chi$ in comoving distance $\chi$ one obtains an expression for the angular displacement potential,
\begin{equation}
\psi = \int\dd\chi\: W_\psi(\chi)\Phi
\quad\mathrm{with}\quad
W_\psi(\chi) = \frac{n(\chi)D_+}{\chi^2},
\end{equation}
which replaces the lensing potential in the case of gravitational lensing of the CMB. The statistical properties of $\psi$, which is a Gaussian random field, are described by the spectrum $C_\psi(\ell)$,
\begin{equation}
C_\psi(\ell) = 
\int\frac{\dd\chi}{\chi^2}\: W_\psi^2(\chi)P_{\Phi}(k=\ell/\chi)
\end{equation}
which results from carrying out a Limber-projection of $\psi$. The spectrum $C_\alpha(\ell)$ is related to $C_\psi(\ell)$ by $C_\alpha(\ell) = l^2C_\psi(\ell)$ as a consequence of the relation $\veca = \nabla_\theta\psi$. 

\begin{figure}
\begin{center}
\resizebox{\hsize}{!}{\includegraphics{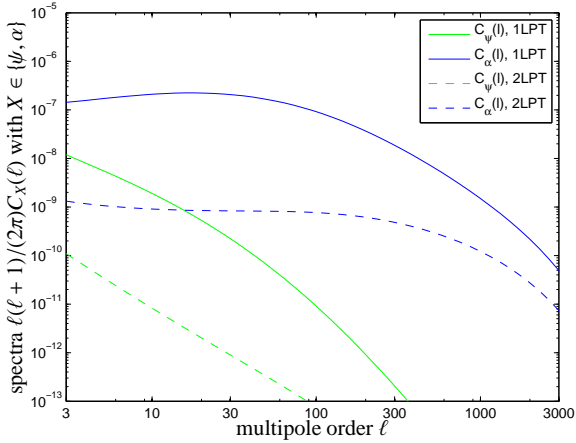}}
\end{center}
\caption{Angular spectrum $C_\psi(\ell)$ (green line) of the displacement potential $\psi$ and the spectrum $C_\alpha(\ell)\equiv l^2C_\psi(\ell)$ (blue line) of the displacement field $\alpha=\nabla_\theta\psi$, for 1LPT (solid line) and 2LPT (dashed line).}
\label{fig_spectrum}
\end{figure}

The angular spectrum $C_\psi(\ell)$ of the displacement potential $\psi$ resulting from the Limber-projection of $P_\Phi(k)$ is depicted in Fig.~\ref{fig_spectrum} along with the spectrum $C_\alpha(\ell) = \ell^2 C_\psi(\ell)$ of the displacement angle $\veca$. Clearly, the 1LPT-result dominates over the 2LPT result by more than one order of magnitude, as already suggested by Fig.~\ref{fig_potential}. The similarity of the plot to the analogous quanities in CMB-lensing is striking.

Correlations between the components of the shifting angle $\alpha$ at two positions $\vect_1$ and $\vect_2$ are described by \citep{1996ApJ...463....1S}
\begin{equation}
\bra\alpha_i(\vect_1)\alpha_j(\vect_2)\ket = \frac{1}{2}C_0(\theta) - C_2(\theta)\:\hat{\theta}_{\langle i}\hat{\theta}_{j\rangle}
\end{equation}
with $\vect=\vect_2-\vect_1$, and correlation functions of the displacement angle which are defined as
\begin{equation}
C_0(\theta) = \int\frac{\ell^3\dd\ell}{2\pi} C_\psi(\ell)J_0(\ell\theta)
\end{equation}
and
\begin{equation}
C_2(\theta) = \int\frac{\ell^3\dd\ell}{2\pi} C_\psi(\ell)J_2(\ell\theta).
\end{equation}
We introduce the abbreviation $\sigma^2(\theta) = C_0(0) - C_0(\theta)$ in complete analogy to CMB-lensing for describing uncorrelated displacements. The characteristic function of a Gaussian displacement field $\veca$ would then be:
\begin{equation}
\left\langle\exp\left(\ci\bmath{\ell}\left[\veca(\vect_1)-\veca(\vect_2)\right]\right)\right\rangle = 
\exp\left(\frac{\ell^2}{2}\left[-\sigma^2(\theta) + \cos2\varphi_\ell C_2(\theta)\right]\right).
\end{equation}
In the case of CMB-lensing, non-Gaussian contributions have been shown to have negligible effect on the deflection angle statistic \citep{2009MNRAS.396..668C, 2011MNRAS.411.1067M} and in the case of weak cosmic shear, analogous arguments about the sparcity of strong deflections apply equally \citep{2005MNRAS.356..829H}.

\begin{figure}
\begin{center}
\resizebox{\hsize}{!}{\includegraphics{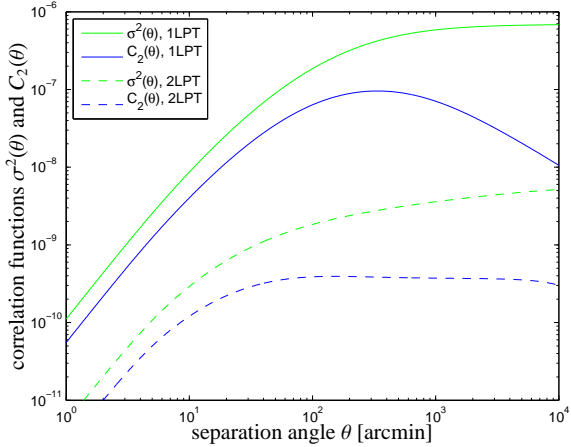}}
\end{center}
\caption{Correlation functions $\sigma^2(\theta) = C_0(0)-C_0(\theta)$ (blue line) and $C_2(\theta)$ (green line) as a function of separation angle $\theta$, for 1LPT (solid line) and 2LPT (dashed line).}
\label{fig_correlation}
\end{figure}

Fig.~\ref{fig_correlation} shows the quantities $\sigma^2(\theta) = C_0(0) - C_0(\theta)$ and $C_2(\theta)$ used in this formalism, for both 1LPT and 2LPT. Again, we would like to draw the reader's attention to the similarity between our results and the formally equivalent result in CMB-lensing and to the domination of the 1LPT results over the 2LPT spectra. 

The correlation properties of the shifted ellipticity field can be described using the two correlation functions $\xi_\pm(\theta)$,
\begin{eqnarray}
\xi_+^\prime(\theta) & = & 
\bra\epsilon^*(\vecx+\veca)\epsilon(\vecx^\prime+\veca^\prime)\ket\\
\xi_-^\prime(\theta) & = & 
\bra\exp(-4\ci\phi_\ell)\epsilon(\vecx+\veca)\epsilon(\vecx^\prime+\veca^\prime)\ket
\end{eqnarray}
where the points at which the ellipticities are observed, are shifted by exactly the angle $\veca$. Substituting the correlation function for the deflection angle in the Fourier-transforms of the above expressions yields the correlation functions $\xi_\pm^\prime(\theta)$ of the shifted ellipticity field. They can be transformed to $E$-mode and $B$-mode spectra with the standard transformation written down in eqns.~(\ref{eqn_e_transform}) and~(\ref{eqn_b_transform}).

In summary, the $E$-mode and $B$-mode spectra of the shifted ellipticity field can be written concisely in a matrix notation:
\begin{equation}
\left(
\begin{array}{c}
C_E^\prime(\ell) \\
C_B^\prime(\ell)
\end{array}
\right) = 
\int\lprime\dd\lprime
\left(
\begin{array}{cc}
W_+(\ell,\lprime) & W_-(\ell,\lprime) \\
W_-(\ell,\lprime) & W_+(\ell,\lprime)
\end{array}
\right)
\left(
\begin{array}{c}
C_E^\epsilon(\lprime)\\
C_B^\epsilon(\lprime)
\end{array}
\right).
\label{eqn_transformation}
\end{equation}
This notation shows explicitly the mixing between scales due to the convolution integral and the conversion between $C_E^\epsilon(\ell)$ and $C_B^\epsilon(\ell)$ under the influence of $W_-(\ell,\lprime)$, which is the non-diagonal entry of the mixing matrix. The kernels $W_\pm(\ell,\lprime$ are given by
\begin{eqnarray}
W_+(\ell,\lprime) & = & \frac{1}{2}\int\theta\dd\theta \left[J_0(\ell\theta)A(\lprime,\theta) + J_4(\ell\theta)B(\lprime,\theta)\right],\\
W_-(\ell,\lprime) & = & \frac{1}{2}\int\theta\dd\theta \left[J_0(\ell\theta)A(\lprime,\theta) - J_4(\ell\theta)B(\lprime,\theta)\right],
\end{eqnarray}
with the functions
\begin{eqnarray}
A(\ell,\theta) & = & 
\exp\left(-\frac{\ell^2\sigma^2(\theta)}{2}\right)\left[J_0(\ell,\theta) + \frac{\ell^2}{2}C_2(\theta)J_4(\ell\theta)\right],\\
B(\ell,\theta) & = &
\exp\left(-\frac{\ell^2\sigma^2(\theta)}{2}\right)\left[J_4(\ell,\theta) + \frac{\ell^2}{2}C_2(\theta)J_s(\ell\theta)\right],
\end{eqnarray}
which describe uncorrelated shifting due to $\sigma^2(\theta)$ and correlated displacements due to $C_2(\theta)$. We abbreviated $J_s(x) = J_2(x) + J_6(x)$. In the limit of no shifting, $C_0(\theta) = C_2(\theta) = 0$ such that $W_+(\ell,\lprime) = \delta(\ell-\lprime) / \ell$ and $W_-(\ell,\lprime) = 0$, due to the orthogonality relations of the cylindrical Bessel functions,
\begin{equation}
\int\theta\dd\theta\: J_n(\ell\theta) J_n(\lprime\theta) = \frac{1}{\ell}\delta_D(\ell-\lprime).
\end{equation}
In this limit, the convolution is reduced to a Dirac $\delta_D$-function and the mixing matrix is the unit matrix, so that the $E$-mode and $B$-mode amplitudes are conserved. We have verified that higher-oder corrections arising in the transformation of correlation functions do have a negligible effect for the evolved ellipticity correlations \citep{2005PhRvD..71j3010C, 2006PhR...429....1L}.

\subsection{$E/B$-mode conversion}

\begin{figure}
\begin{center}
\resizebox{0.9\hsize}{!}{\includegraphics{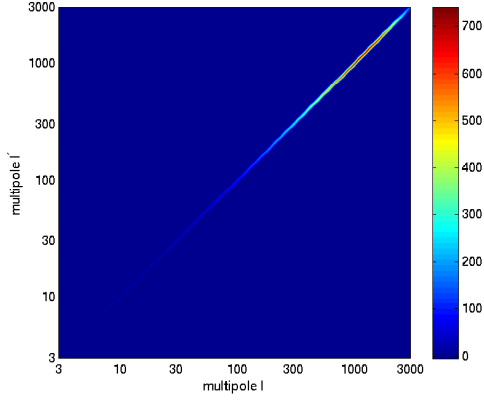}}
\end{center}
\caption{Mode coupling kernel $(\ell\lprime)\times W_+(\ell,\lprime)$ used in the transformation of the ellipticity spectra. For simplicity, we focus on 1LPT because the contributions due to 2LPT are small.}
\label{fig_kernel_plus}
\end{figure}

\begin{figure}
\begin{center}
\resizebox{0.9\hsize}{!}{\includegraphics{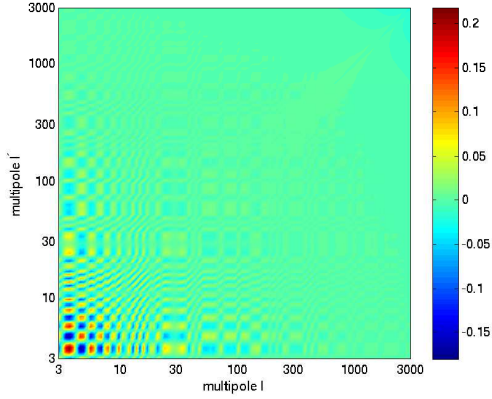}}
\end{center}
\caption{Mode coupling kernel $(\ell\lprime)\times W_-(\ell,\lprime)$ responsible for the $E\leftrightharpoons B$-mode conversion in the ellipticity field. Again, we show the results for 1LPT because the contributions from 2LPT are small.}
\label{fig_kernel_minus}
\end{figure}

Figs.~\ref{fig_kernel_plus} and~\ref{fig_kernel_minus} show the mode coupling kernels $W_+(\ell,\lprime)$ and $W_-(\ell,\lprime)$ where for simplicity we focus on 1LPT because the contributions due to 2LPT are comparatively small. From Fig.~\ref{fig_kernel_plus}  we see that the power of the $W_+(\ell,\lprime)$-kernel is mainly distributed along the diagonal and increasing with multipole number, with maximum contribution from $300\lsim\ell\lsim3000$. The off-diagonal contribution creates a convolution (eqn.~\ref{eqn_transformation}) between the spectra at different multipoles, mediated by $W_+(\ell,\lprime)$. In contrast, the mode coupling kernel $W_-(\ell,\lprime)$ (Fig.~\ref{fig_kernel_minus}), which is responsible for the $E$/$B$-conversion, shows a lateral pattern which is three orders of magnitude smaller in amplitude and decreasing with higher multipole numbers $(\ell,\ell')$.

\subsection{Ellipticity spectra}
\begin{figure}
\begin{center}
\resizebox{\hsize}{!}{\includegraphics{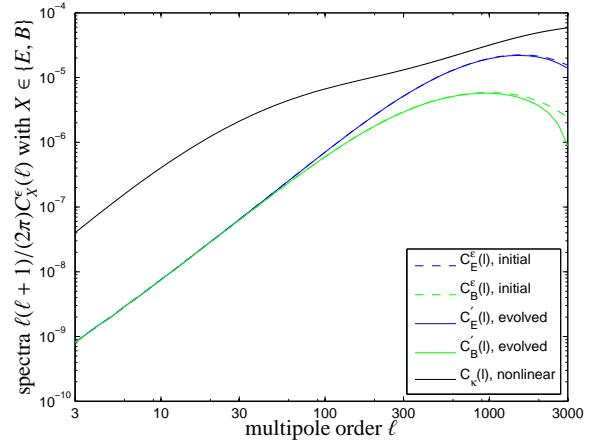}}
\end{center}
\caption{Ellipticity spectra $C_E^\epsilon(\ell)$ (blue line) and $C_B^\epsilon(\ell)$ (green line) as predicted by the angular momentum model with $a=0.25$ and the disk thickness parameter set to $\alpha=1$ (dashed line), and the evolved ellipticity spectra (solid line) where the displacements were computed by 1LPT. For comparison, we plot the spectrum $C_\kappa(\ell)$ of the weak lensing convergence for a linear (black dashed line) and nonlinear (black solid line) CDM spectrum.}
\label{fig_ellipticity}
\end{figure}

The final result is given in Fig.~\ref{fig_ellipticity}, which compares the initial ellipticity spectra $C_E^\epsilon(\ell)$ and $C_B^\epsilon(\ell)$ of the ellipticity field as predicted by the angular momentum model, and the evolved spectra $C_E^\prime(\ell)$ and $C_B^\prime(\ell)$ due to peculiar motion.
For comparison with weak lensing, we plot the weak convergence spectrum $C_\kappa(\ell)$ expected from the EUCLID galaxy sample in comparison, for a nonlinear CDM spectrum \citep[using the parameterisation by][]{2003MNRAS.341.1311S}. The first observation is that ellipticity correlations reach amplitudes similat to those of the weak lensing convergence in the nonlinear part corresponding to amplitudes $\ell\lsim300$, and that the intrinsic $E$-mode spectrum $C_E^\epsilon(\ell)$ is larger than the $B$-mode spectrum $C_B^\epsilon(\ell)$ by about an order of magnitude in this regime. On larger angular scales, there are no appreciable ellipticity correlations and one effectively observes the variance of the ellipticity field for uncorrelated objects. Consequently, the spectra have identical amplitudes and are effectively constant. In this regime, the shifting effect is not able to affect the galaxies, which is a well-known result in CMB-lensing, where scale free-spectra are invariant \citep{2006PhR...429....1L}: The mode-conversion mechanism is uneffective if the spectra are equal, $C_E^\epsilon(\ell) = C_B^\epsilon(\ell)$, and the convolution with $W_+(\ell,\lprime)$ is not able to redistribute amplitudes. In contrast, both spectra are affected on multipoles $\ell>1000$, where in particular $C_B^\prime(\ell)$ has decreased relative to $C_B^\epsilon(\ell)$.

\begin{figure}
\begin{center}
\resizebox{\hsize}{!}{\includegraphics{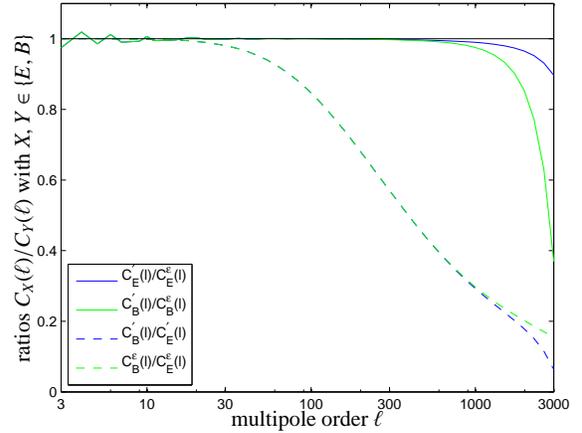}}
\end{center}
\caption{Ratios $C^\prime_E(\ell)/C^\epsilon_E(\ell)$ (blue solid line), $C^\prime_B(\ell)/C^\epsilon_B(\ell)$ (green solid line), $C^\prime_B(\ell)/C^\prime_E(\ell)$ (blue dashed line) and $C^\epsilon_B(\ell)/C^\epsilon_E(\ell)$ (green dashed line) with all displacements following from 1LPT.}
\label{fig_ratios}
\end{figure}

Fig.~\ref{fig_ratios} compares the relative magnitude of all spectra as a function of multipole $\ell$. The plot shows the  relative ratio of the evolved and initial $E$-mode and $B$-mode spectra. As already indicated by Fig.~\ref{fig_ellipticity}, we see a significant decrease for $l>1000$ of up to 10\% for the $E$- and 60\% for the $B$-modes at $l\simeq 3000$. The ratios $C_B^\prime(\ell)/C_E^\prime(\ell)$ and $C_B^\epsilon(\ell)/C_E^\epsilon(\ell)$ of intrinsic and evolved spectra are similar up to multipoles of $\ell\simeq1000$, where they separate and indicate that the newly generated $B$-modes are small and that the $B$-mode spectra are more strongly affected. For EUCLID's weak lensing application, changes in the ellipticity spectra are affecting scales where the shape noise starts dominating, but for shallower surveys, lower multipoles would be affected by the peculiar motion effect.

\begin{figure}
\begin{center}
\resizebox{\hsize}{!}{\includegraphics{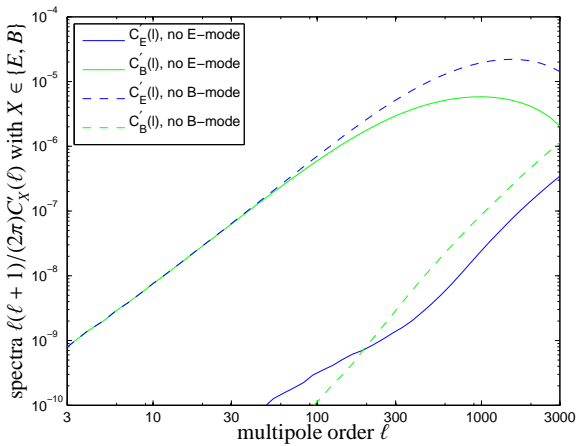}}
\end{center}
\caption{Contributions to the evolved ellipticity spectra $C_E^\prime(\ell)$ (blue lines) and $C_B^\prime(\ell)$ (green lines): no initial $B$-mode spectrum, $C_B^\epsilon(\ell) = 0$ (dashed lines) and no initial $E$-mode spectrum, $C_E^\epsilon(\ell)=0$ (solid lines).}
\label{fig_contrib}
\end{figure}

Finally, Fig.~\ref{fig_contrib} gives an impression of the mode conversion mechanism, where we plot evolved spectra $C_E^\prime(\ell)$ and $C_B^\prime(\ell)$, when the $E$-mode or the $B$-mode in the initial spectra was deliberately set to zero, i.e. $C_E^\epsilon(\ell)=0$ in the first and $C_B^\epsilon(\ell)=0$ in the second case. Even in the absence of a particular initial mode we observe power in the corresponding evolved spectrum, as a consequence of $E$/$B$-coupling introduced by peculiar motion.

\section{Summary}\label{sect_summary}
The topic of this paper is the evolution of intrinsic ellipticity correlation between galaxies due to peculiar motion. Intrinsic ellipticity correlations are derived in the framework of angular momentum models, which explain these correlations by correlated tidal shears experienced by the protohaloes in acquiring their angular momenta. Because the symmetry axis of the galactic disk is related to the angular momentum direction of the host halo, correlated angular momenta give rise to correlated angles of inclination and hence correlated ellipticities. 

\begin{enumerate}
\item{Peculiar motion of galaxies changes the correlation properties of the ellipticity field by displacing the galaxies and distorting the ellipticity field. We describe the peculiar motion by Lagrangian perturbation theory and derive corresponding displacement angles along with their statistical properties for the EUCLID galaxy sample. The formalism for evolving the ellipticity spectra uses an analogy to the formalism describing lensing of the CMB polarisation spectra. Both quantities, the ellipticity field as well as the polarisation field, have the same symmetry properties, being of spin 2. The loci at which ellipticities and polarisations are measured are displaced by peculiar motion in the first and by gravitational lensing in the second case. Because the peculiar motion field in the quasi-linear regime is a flow resulting from a velocity potential which corresponds to the lensing potential, is it possible to derive all necessary quanities in complete analogy.}
\item{Peculiar motion has two effects on the ellipticity spectra: There is a convolution of the spectra and a conversion between $E$-modes and $B$-modes of the ellipticity field. Both effects become important on angular scales $\ell>1000$, because on smaller multipoles, the spectra are effectively constant and equally large. In particular the spectrum $C_B^\epsilon(\ell)$ is strongly affected and looses amplitude: For the EUCLID galaxy sample we measure decrements by about 10\% for $C_E^\epsilon(\ell)$ and 60\% for $C_B^\epsilon(\ell)$. The mode-conversion mechanism is comparatively weak and we tested it by deliberately setting the initial spectra $C_E^\epsilon(\ell)$ and $C_B^\epsilon(\ell)$ to zero.}
\item{Second order corrections in the dynamical model were found to be negligibly small in comparison to first order Lagrangian perturbation theory. Likewise, we made sure that higher-order corrections in the transformation of the ellipticity spectra had a minor effect on the evolved ellipticity spectra.}
\end{enumerate}
We conclude that in principle the dispersing effect of peculiar motion weakens intrinsic ellipticity correlations and make them less troublesome for analysing weak lensing data. For the case of EUCLID we see changes in the spectra on scales where the shape noise is already dominating. A natural extention to this investigation would comprise the shifting and distorting effect of weak gravitational lensing, and ultimately the usage of analysis methods conceived for the polarisation of the CMB for investigating intrinsic ellipticity correlations.

\section*{Acknowledgements}
Our work was supported by the German Research Foundation (DFG) within the framework of the excellence initiative through the Heidelberg Graduate School of Fundamental Physics. In particular, we acknowledge funding from the FRONTIER-programme. We would like to thank Philipp M. Merkel for his suggestions and advice on numerical computations.

\bibliography{bibtex/aamnem,bibtex/references}
\bibliographystyle{mn2e}

\appendix

\bsp

\label{lastpage}

\end{document}